# SPIDER ION SOURCE AND EXTRACTION POWER SUPPLIES - AN UPDATE OF THE DESIGN OF THE BIAS CIRCUITS AFTER FOUR YEARS OF OPERATION

Alastair Shepherd[1,2], Marco Bigi[ξ,1], Riccardo Casagrande[1], Mattia Dan[1], Alberto Maistrello[1], Emanuele Sartori[1,3,4], Gianluigi Serianni[1,5], Hans Decamps[6], Loris Zanotto[1]

[1] *Consorzio RFX (CNR, ENEA, INFN, Università di Padova, Acciaierie Venete SpA ), C.so Stati Uniti 4, 35127 Padova, Italy*
[2] *CCFE, Culham Science Centre, Abingdon, Oxfordshire OX14 3DB, UK*
[3] *Department of Management and Engineering, Università degli Studi di Padova, Strad. S. Nicola 3, 36100 Vicenza, Italy*
[4] *CRF – University of Padova, Italy*
[5] *ISTP-CNR, Padova, Italy*
[6] *ITER Organization, Route de Vinon-sur-Verdon, CS 90 046, 13067 St Paul Lez Durance Cedex - France*

*Abstract*

SPIDER, the ion source prototype for ITER neutral beam injectors, has been in operation since 2018. The experiment includes a system of Ion Source and Extraction Power Supplies (ISEPS), whose design dates back to 2011. Since then the experience gained has led to substantial changes, the most dramatic modification being the complete replacement of the radiofrequency power supply, described elsewhere. This contribution details operating experience and calculations that have led to defining new requirements for a specific subset of power supplies, the bias power supplies.

The bias power supplies perform the function of filtering out electrons extracted together with the negative ions and in the original 2011 ISEPS design are single quadrant resonant converters. For the two bias power supplies, the picture resulting from the review described in the paper is a substantial redesign, with the ability to operate in four quadrants and a twofold increase of the rated output voltage the most significant new specifications.

## I. INTRODUCTION

The ITER Neutral Beam Test Facility in Padua (Italy), hosts SPIDER [1,2], the full scale prototype of the ion source for ITER Neutral Beam Injectors. In the ion source, plasma is produced by inductive coupling of radiofrequency power inside eight cylindrical regions knowns as "drivers". From the electrical point of view, the drivers are arranged in pairs, each pair being fed by a radiofrequency generator providing 200 kW cw power at 1 MHz frequency. Electrostatic extraction and acceleration of a negative ion beam takes place in a three-electrode system, with the electrode closest to the ion source termed plasma grid (PG). Short-circuits between the grids of the extractor and accelerator are a common operational feature of neutral beam systems.

The Ion Source and Extraction Power Supplies (ISEPS) [3], provide power to the ion source for production and extraction of the negative ions. The design of the ISEPS of SPIDER dates back to 2010-2011 and was based on a collection of requirements and ratings within the neutral beam community, with inputs from the operating experience of negative ion sources in particular [4]. The ISEPS of SPIDER were manufactured between 2012 and 2013, factory tested in 2014, installed in 2015 and site tested in 2016. At that point, the system was handed over by the industrial supplier to the SPIDER operation team and in 2017 the one-to-one integration with the experiment control system took place [5]. Finally, in June 2018 SPIDER operation with plasma commenced [6].

## II. SPIDER BIAS CIRCUITS

Experiments on negative ion test beds showed that positive polarisation of the plasma grid with respect to the body of the ion source would reduce the number of co-extracted electrons. The so-called bias plate (BP), a metallic structure in front of the plasma grid that is also positively biased, enhances the preferential drain of electrons from the source plasma further [7]. The design of SPIDER includes two biasing circuits, as shown in Figure 1. In the first circuit, a power supply (labelled ISBI) applies a positive voltage to the PG with respect to the ion source body. In the second circuit, a separate power supply (labelled ISBP) applies a positive voltage to the BP electrode with respect to the ion source body. Both power supplies, ISBI and ISBP are part of the ISEPS system. The output ratings of ISBI and ISBP in the 2011 ISEPS design are given in Table 1.

---

ξ email: marco.bigi@igi.cnr.it

**Table 1.** Specifications of SPIDER bias power supplies as of ISEPS 2011 design.

| Power supply | Output ratings |
| --- | --- |
| Source Bias (ISBI) | 30 V   600 A |
| Bias Plate (ISBP) | 30 V   150 A |

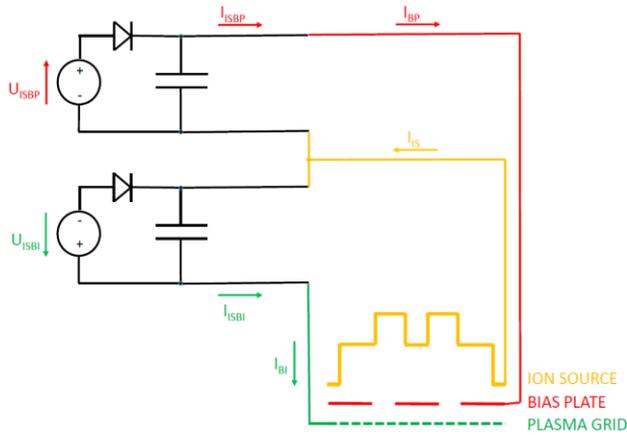

**Figure 1.** Configuration of SPIDER bias circuits at the start of SPIDER operation.

The structure of the ISBI and ISBP power supplies of the original ISEPS design is based on a power module with double resonant architecture, a conceptual circuit diagram of which is given in Figure 2. The power supplies operate on a single quadrant and have no ability to receive steady state power from the load, the output stage being composed of a filter capacitance. Downstream of the power module transformer there are no active components, which is a desirable feature from the point of view of resilience to grid breakdowns [8].

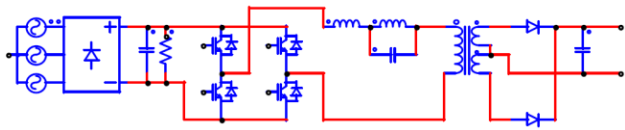

**Figure 2.** Conceptual circuit diagram of the ISBI/ISBP power module.

### III. ISSUES WITH THE ORIGINAL DESIGN OF SPIDER BIAS CIRCUITS

The start of SPIDER operation revealed right away a showstopper associated to the bias circuits [6,9]. In the configuration of Figure 1, there is no dc path for energy feedback from the plasma. Upon plasma ignition, this would cause overvoltage trips of ISBI, ISBP and other source support power supplies. Similar trips would be recorded during breakdowns of the radiofrequency circuit, which were comparatively frequent at the beginning of SPIDER operation [6]. Measures were subsequently taken to reduce the gas pressure at the back of the source, with the installation of a mask reducing gas conductance of the PG [10,11].

A second problem was that in caesium free operation, even at a moderate radiofrequency power of 50 kW/driver, the plasma potential would be higher than the bias power supply output voltage rating (30 V) [12]. This is seen in Figure 3 where the output voltages $U_{ISBI}$ and $U_{ISBP}$ of the two power supplies are above 30 V at PG currents in the range 0.5-2.8 kA with ISBI and ISBP disabled (potentials defined by the plasma, zero output current $I_{ISBI/BP}$). Consequently, it would not be possible to employ the bias circuits for their prime function of modifying the potential distribution around the grids and reducing co-extracted electrons. Furthermore, at higher RF power and with all four radiofrequency generators in operation, it would be likely to exceed the overvoltage threshold of ISBI and ISBP power supplies.

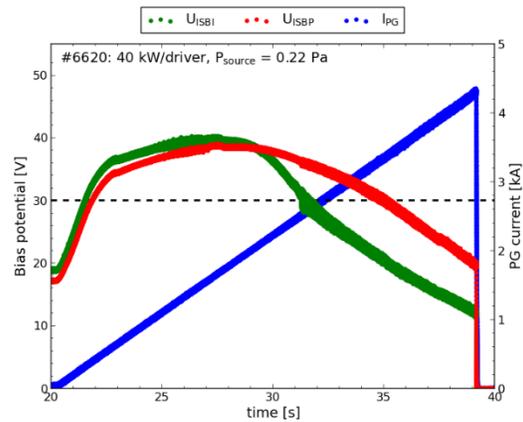

**Figure 3.** SPIDER pulse 6620, showing variation of 'floating' PG and BP bias potentials with varying PG current. With 0.6 Ω parallel output resistors as explained in section IV.

### IV. CHANGES MADE TO SPIDER BIAS CIRCUITS IN 2018/2019

In response to the overvoltage trips described above, a first modification of the bias circuit was carried out in the very early stages of SPIDER operation in 2018. A 0.6 Ω resistor was added in parallel at the output of ISBI and ISBP power supplies (see Figure 4). In this new configuration, stable plasma ignition and operation were achieved [6,9].

A second change (summer 2019) consisted in replacing the ISBI and ISBP power modules with ones having higher output voltage rating, while keeping the 0.6 Ω parallel resistors installed the previous year. The new ratings are summarised in Table 2. Once the power supplies were replaced, the bias potentials could be controlled when the plasma defined potential was above 30 V (Figure 5, pulse #6652, potential defined by plasma before 25 s, potential controlled by power supplies after 25 s).

**Table 2.** Output ratings of the power modules employed on SPIDER bias power supplies, as of summer 2019.

| Power supply | Output ratings |
|---|---|
| Source Bias (ISBI) | 50 V   200 A |
| Bias Plate (ISBP) | 50 V   200 A |

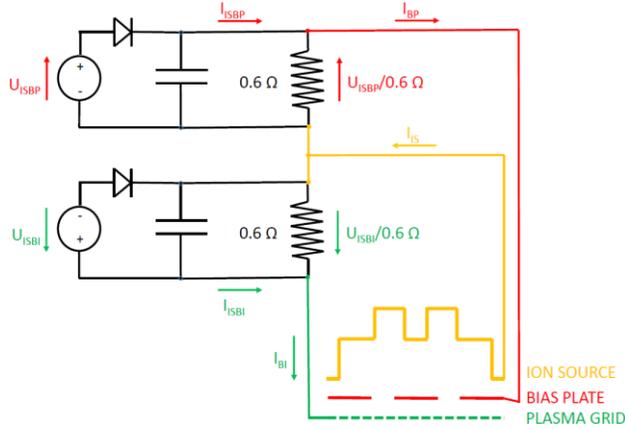

**Figure 4.** Modified and current configuration of SPIDER bias circuits.

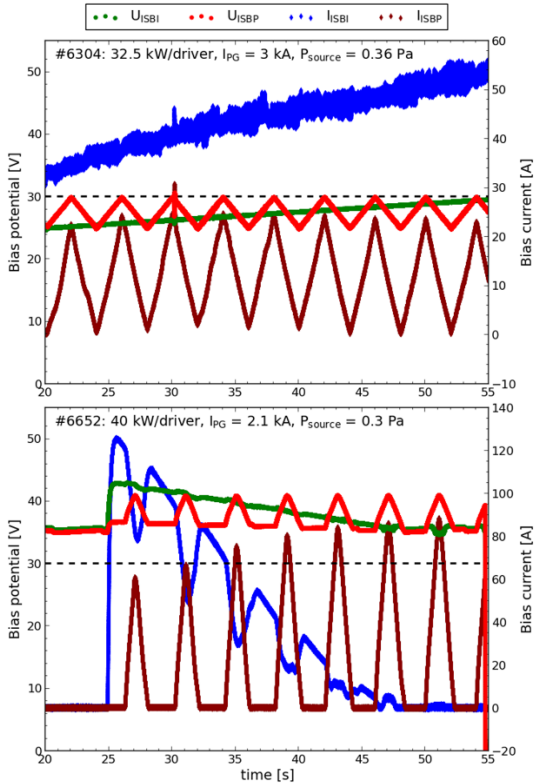

**Figure 5.** Bias and bias plate voltage scans. Pulse #6304 used the original 30 V power supplies and 0.42 Ω parallel output resistors. Lower RF power and higher PG current were required to reduce the plasma defined potential below 30 V and allow voltage control. Pulse #6652 used the replacement 50 V power supplies and 0.6 Ω resistors.

## V.    2021 BIAS OPERATION

During 2021 SPIDER operated with evaporated caesium in the source for the first time [2], to increase the production of negative ions and reduce the co-extracted electron current. During the campaign, S21, the ISBI and ISBP power supplies were primarily operated in current control, where the control system controls the output currents $I_{ISBI}$ and $I_{SBP}$. Maximum bias currents of $I_{ISBI}$ = 190 A and $I_{ISBP}$ = 140 A were used, with ISBI limited by the power supply and ISBP limited by the 185 A rating of its feedthrough, further limited to 140 A as a safety margin.

The bias voltages $U_{ISBI}$ and $U_{ISBP}$ and the net currents to the plasma $I_{BI}$ and $I_{BP}$ are given in Figure 6 and Figure 7. The net currents provided by the power supplies to the plasma, due to the presence of the resistors in parallel to the power supply outputs as shown in Figure 4, are given by equations (1) and (2). Discounting the effect of the plasma impedance and the extracted current (< 2 A with caesium and a reduced number of apertures [2]), the return current $I_{IS}$ is given by (3).

$$I_{BI} = I_{ISBI} - \frac{U_{ISBI}}{R_{BI}} \quad (1)$$

$$I_{BP} = I_{ISBP} - \frac{U_{ISBP}}{R_{BI}} \quad (2)$$

$$I_{IS} = I_{BI} + I_{BP} \quad (3)$$

In the pulses before caesiation the bias voltage reached 45 V, with the bias plate 1-2 V lower, as is usual for SPIDER when the ISBI and ISBP currents are set equally. With the introduction of Cs, the bias and bias plate potentials decrease by several volts. The larger potential difference between the PG and bias plate is due to $I_{ISBI}$ initially being controlled to 80 A and $I_{ISBP}$ to 0 A. Later when controlling ISBI and ISBP to the same current the potential difference decreases to 1-2 V. During Cs operation, the ISBI voltage reaches a maximum of 43 V, with net currents $I_{BI}$ and $I_{BP}$ of 135 A and 88 A.

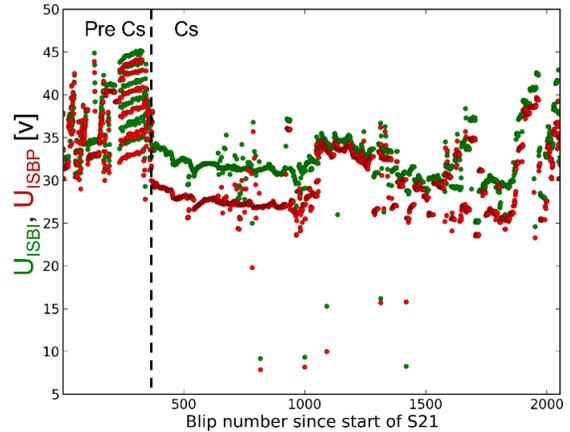

**Figure 6.** Bias voltages during SPIDER operation in 2021. Each SPIDER pulse can be up to one hour long, containing repeated plasma phases called "blips".

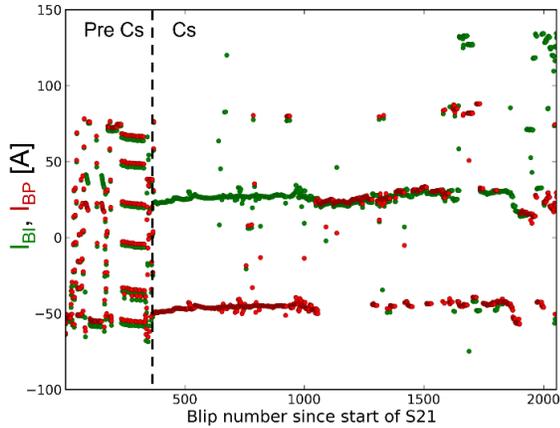

**Figure 7.** Net bias currents during SPIDER operation in 2021.

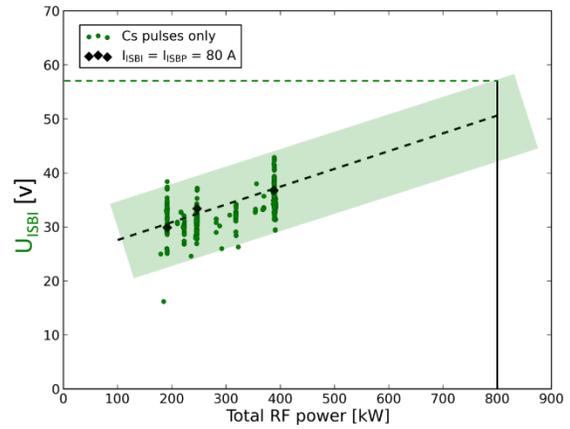

**Figure 8.** Extrapolation of ISBI voltage to 800 kW total RF power (100 kW/driver). The green projections are to lead the eye and provide an estimate. Black points are at fixed bias current $I_{ISBI} = I_{ISBP} = 80$ A and similar source parameters, to show that the projection is feasible.

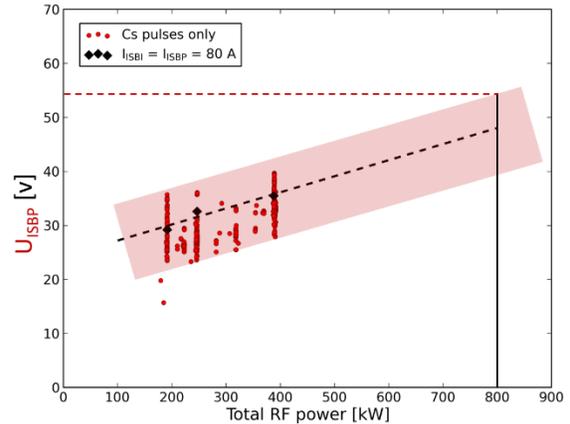

**Figure 9.** Extrapolation of ISBP voltage to 800 kW total RF power (100 kW/driver). The red projections are to lead the eye and provide an estimate. Black points are at fixed bias current $I_{ISBI} = I_{ISBP} = 80$ A and similar source parameters, to show that the projection is feasible.

## VI. HIGH POWER EXTRAPOLATION

From a physics perspective, the net bias currents, being the current of electrons (or ions if negative) drained from the plasma, are of more interest. For future operation, with the introduction of solid-state amplifiers, the RF power provided to the plasma will increase from a maximum of 50 kW/driver to 100 kW/driver. With increased RF power the plasma density can be expected to increase, and with it the plasma potential and net current required to reduce the number of co-extracted electrons.

Plotting the bias voltages against RF power for the Cs pulses in Figure 8 and Figure 9, the bias and bias plate voltages show a general increase with RF power. As a variety of source parameters can influence the plasma, and thus the bias voltages, there is a large range at fixed RF power. Pulses at different RF powers with the bias and bias plate currents controlled to 80 A show the general trend of increasing bias voltage. At full RF power the ISBI voltage is expected to exceed 55 V, with the ISBP voltage 1-2 V lower. This exceeds the present 50 V rating of the ISBI and ISBP power supplies.

Taking the net bias currents for ISBI and ISBP against the bias voltages (Figure 10 and Figure 11), it is possible to extrapolate the net electron current drained at 100 kW/driver. The PG bias current at 57 V bias voltage reaches 600 A, the rating of the original ISBI power supply. The bias plate current also reaches above 500 A.

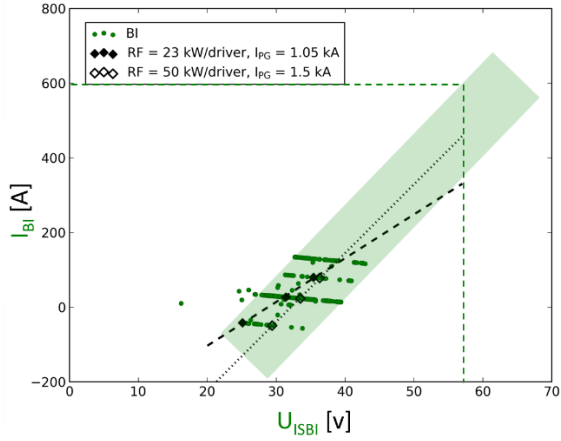

**Figure 10.** Extrapolation of ISBI net current to ISBI voltage at 800 kW. The green projections are to lead the eye and provide an estimate. Black points are from bias current scans at two RF powers to show that the projection is feasible.

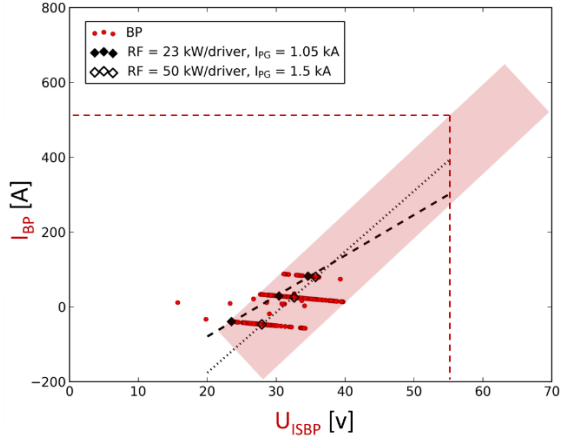

**Figure 11.** Extrapolation of ISBP net current to ISBP voltage at 800 kW. The red projections are to lead the eye and provide an estimate. Black points are from bias current scans at two RF powers to show that the projection is feasible.

## VII. NEGATIVE BIAS CURRENT

The resistors in parallel to the bias power supplies provide a path for energy feedback from the plasma. In these cases, such as when the PS is controlled to 0 A so that the voltage is defined by the plasma, a negative net current ($I_{BI/BP}$) flows through the resistors. This corresponds to ions being drained from the plasma. To widen the scientific window of SPIDER it is desirable to increase the negative current capability of the bias and bias plate power supplies. This would allow the plasma electrodes to be moved below the 'floating' voltage towards the operational conditions of MITICA bias plate, which will be shorted to the source wall, without doing something similar on SPIDER.

While the bias surfaces cannot be considered a true Langmuir probe, as the surface is not infinitesimally small compared to the plasma and drawing the current perturbs the plasma, the negative current required by the power supply can be roughly estimated by the positive ion saturation current (equation 4). For a plasma with an electron temperature $T_e$ = 2 eV and density $n_i$ = 2x10$^{17}$ m$^{-3}$ the surface of the PG and BP (~2 m$^2$) has an ion saturation current in the range of 100s of amps.

$$j_i \sim q_e n_i \sqrt{\frac{k_B T_e}{m_i}} \qquad (4)$$

## VIII. NEW REQUIREMENTS FOR SPIDER BIAS POWER SUPPLIES

The present configuration and ratings of SPIDER bias circuits have a number of implications and drawbacks:
- 50 V output rating is insufficient for operation at full radiofrequency power (100 kW per driver), even in the presence of caesium,
- 200 A ISBI current rating is insufficient for operation at nominal extracted current, i.e. once the Plasma Grid mask [10] is removed,
- The parallel resistor (see Fig. 2) draws part of the power supply output current, effectively reducing the amount of current available for the true biasing function,
- The power supply cannot output negative voltage.

To overcome the limitations, based on the data and extrapolations shown above, requirements have been identified for new bias power supplies, see Table 3. With respect to the original design of Table 1, the most significant changes are the ability to operate on four quadrants and the two-fold increase of the output voltage. 2021 data showed no need to modify the current rating of ISBI compared to the initial rating of 600 A; on the contrary, a significant increase of ISBP current capability would have been desirable as the BP has larger surface for draining electrons and standard operation is with a small difference between ISBI and ISBP bias potentials. However, due to the ISBP downstream vacuum feedthrough current rating (185 A) and the ion source busbar and feedthrough current rating (750 A for ISBI+ISBP) the requirements in Table 3 cannot be exceeded. The constraint on ISBP current will result in some limitations in the operations; these can be overcome in a future shutdown, by replacing busbars and feedthroughs, after a more precise experimental verification of the requirements in the next experimental operations at high RF power. It is also worth mentioning that the effect of ISBP on the beam features is lower than ISBI [12], so that such temporary limitation is deemed tolerable. The -10 V rating is to allow the bias potential to go near 0 V without the risk of tripping the power supply.

Table 3. Requirements for new SPIDER bias power supplies.

| DC output voltage range | -10 / +70 V |
|---|---|
| Rated output current | ±600 A (I_BI) <br> ±200 A (I_BP) |
| Output voltage max ripple | ±200 mV * |
| Output current max ripple | ±3 A * |
| Switch-off time | 20 ms |
| Rise time (10 – 90%) of the output voltage | 100 ms |

* As measured on resistive test load

A preliminary design of new bias power supplies, based on a four-quadrant converter is being assessed. The critical aspects of the design have to do with dissipation of power received from the plasma, redefinition of the output filter and resilience to grid breakdowns. With respect to the latter point, inclusion of actively commanded devices (e.g. Insulated Gate Bipolar Transistors) without downstream galvanic insulation would make the design weaker than the original one.

## IX. CONCLUSIONS AND FUTURE WORK

Temporary modifications and new requirements of SPIDER bias power supplies have been driven by experimentation, since higher current and voltage are needed at higher RF power.

The next phase of this work will consist of designing with an external company an electrotechnical scheme for the ISBI and ISBP power supplies, to achieve the desired specifications. Another input to the design will come from the experience gained in increasing the immunity of the ISEPS system to the effects of grid breakdowns [13].

## X. ACKNOWLEDGEMNT

This work has been carried out within the framework of the ITER-RFX Neutral Beam Testing Facility (NBTF) Agreement and has received funding from the ITER Organization. The views and opinions expressed herein do not necessarily reflect those of the ITER Organization.

This work has been carried out within the framework of the EUROfusion Consortium, funded by the European Union via the Euratom Research and Training Programme (Grant Agreement No 101052200 — EUROfusion). Views and opinions expressed are however those of the author(s) only and do not necessarily reflect those of the European Union or the European Commission. Neither the European Union nor the European Commission can be held responsible for them.